\newcommand{\CLASSINPUTbaselinestretch}{1.61}
\documentclass[12pt,onecolumn,draftclsnofoot,journal,twoside]{IEEEtran}
\ifCLASSINFOpdf
\else
\fi
\usepackage{xfrac}
\usepackage{amsmath}
\usepackage{multirow}
\usepackage{color}
\usepackage{amssymb}
\DeclareMathOperator*{\argmin}{argmin}
\usepackage[belowskip=-20pt,aboveskip=7pt]{caption}
\begin{document}
%
\title{Spatial Multiple Access (SMA): Enhancing performances of MIMO-NOMA systems}
%
\author{Ferdi KARA,~\IEEEmembership{Student Member~IEEE,}
        Hakan KAYA
\thanks{ This work is supported by the Scientific and Technological Research Council of Turkey (TUBITAK) under the 2211-E program. }
\thanks{F. Kara and H. Kaya is with the Department of Electrical and Electronics Engineering, at Zonguldak Bulent Ecevit University. (email: \{f.kara, hakan.kaya\}@beun.edu.tr) }
}
%
%

\markboth{}%
{Kara \MakeLowercase{and} Kaya: Spatial Multiple Access (SMA): Enhancing performances of MIMO-NOMA systems}
%
\IEEEaftertitletext{\vspace{-2.8\baselineskip}}
\maketitle
\begin{abstract}
The error performance of the Non-Orthogonal Multiple Access (NOMA) technique suffers from the inter-user interference (IUI) although it is a promising technique for the future wireless systems in terms of the achievable sum rate. Hence, a multiple access technique design with limited IUI and competitive to NOMA in terms of spectral efficiency is essential.  In this letter, we consider so-called spatial multiple access (SMA) which is based on applying the principle of spatial modulation (SM) through the different users' data streams, as a strong alternative to MIMO-NOMA systems. The analytical expressions of bit error probability (BEP), ergodic sum rate and outage probability are derived for the SMA. The derivations are validated via computer simulations. In addition, the comparison of the SMA system with NOMA is presented. The results reveal that SMA outperforms to the NOMA in terms of the all performance metrics (i.e., bit error rate (BER), outage probability and ergodic sum rate) besides it provides low implementation complexity.
\end{abstract}
%
\IEEEpeerreviewmaketitle
\section{Introduction}
%
%
%
%
\IEEEPARstart{N}{on}-ORTHOGONAL Multiple Access (NOMA) technique is seen as one of the strongest candidates for the future wireless systems [1]. NOMA principle allows serving multiple users at the same resource blocks by splitting them into power domain so that NOMA outperforms to orthogonal multiple access (OMA) techniques in terms of achievable sum rate and the outage probability [2]. This is achieved by implementing superposition coding (SC) at the transmitter and successive interference canceler (SIC) at the receivers [3]. NOMA has tremendous recent attention from the researchers due to its potential. However, most of these studies assume serving only two NOMA users since increasing number of NOMA users boosts system complexity as well as it limits the advantage of NOMA due to the inter-user-interference (IUI). In addition, the authors in [4] showed that the error performance of the NOMA cannot compete with the OMA systems for the both users even if only two NOMA users are served. Hence, the trade-off between the gain in the outage and capacity performances and the decay in the bit error performance caused by IUI is very questionable. Moreover, by considering the complexity at the receiver to implement SIC process, NOMA may not be a feasible solution when two users are served.


Spatial modulation (SM) is another technique proposed for spectral efficiency in MIMO systems [5]. In SM, modulation is held by splitting the input data stream into two groups. While one of the groups is modulated by a M-ary modulation scheme, the other group determines which transmitting antenna will be activated. 
Then, the space shift keying (SSK) is proposed as a subset of SM where the input data stream is transmitted by only mapping to transmitting antenna selection [6]. Multi-user (MU) SM schemes are investigated in the literature but mostly for the uplink scenario [7]. In [8], the authors analyzed the performance of the MU-SM with a channel precoding at the transmitter in a downlink scenario. Although SM/SSK is a spectral efficient technique, the MU applications, in which all users are served by SM/SSK, boost the system complexity due to channel precoder and the need of full channel state information at transmitter (CSIT) so that make them impractical. 

There are also some studies in the literature which consider NOMA and SM principles together [9-10]. Nevertheless, these applications still encounter IUI so the SIC is needed at the receivers. Hence, the low error performance and the implementation complexity is still undergone. In [11] authors point out the challenges of the NOMA networks and consider SM assisted MIMO-NOMA networks and simulations results are provided. However, the analytical analyses are not regarded.

In this letter we analyze the spatial multiple access (SMA) technique  which is is based on implementing SM principle for the input data streams of the different users, for MIMO systems. SMA allocate users into different domains (i.e., spatial and power) rather than only power domain as in NOMA so that the users meet IUI free communication. Hence, SMA achieves the error performance of the OMA systems in addition to providing better outage and capacity performances than conventional NOMA systems. SMA activates only one transmitting antenna during one symbol duration so that the needed radio frequency (RF) chain number is limited to only one. Moreover, not needing SIC implementation at the receivers provides less complexity and latency than NOMA. The rest of paper is organized as follows. In section II, the SMA system model is introduced and maximum likelihood (ML) detections at the users are given. In section III, the performance analyses of the SMA system are given in terms of the bit error probability, capacity and the outage probability. Then, the validation of the derived expressions via computer simulations are presented in addition to the simulation comparisons of the SMA and NOMA. Finally, in section IV the results are discussed and the paper is concluded.

\section{System Model}
We consider a downlink MIMO scenario where a base station (BS) and two mobile users (i.e., UE-1 and UE-2) are located. BS is equipped with $N_t$ antennas whereas each user is equipped with $N_r$ antennas. The spatial multiple access system model is shown in Fig1.
\begin{figure}[!t]
    \centering
    \includegraphics[width=8.5cm]{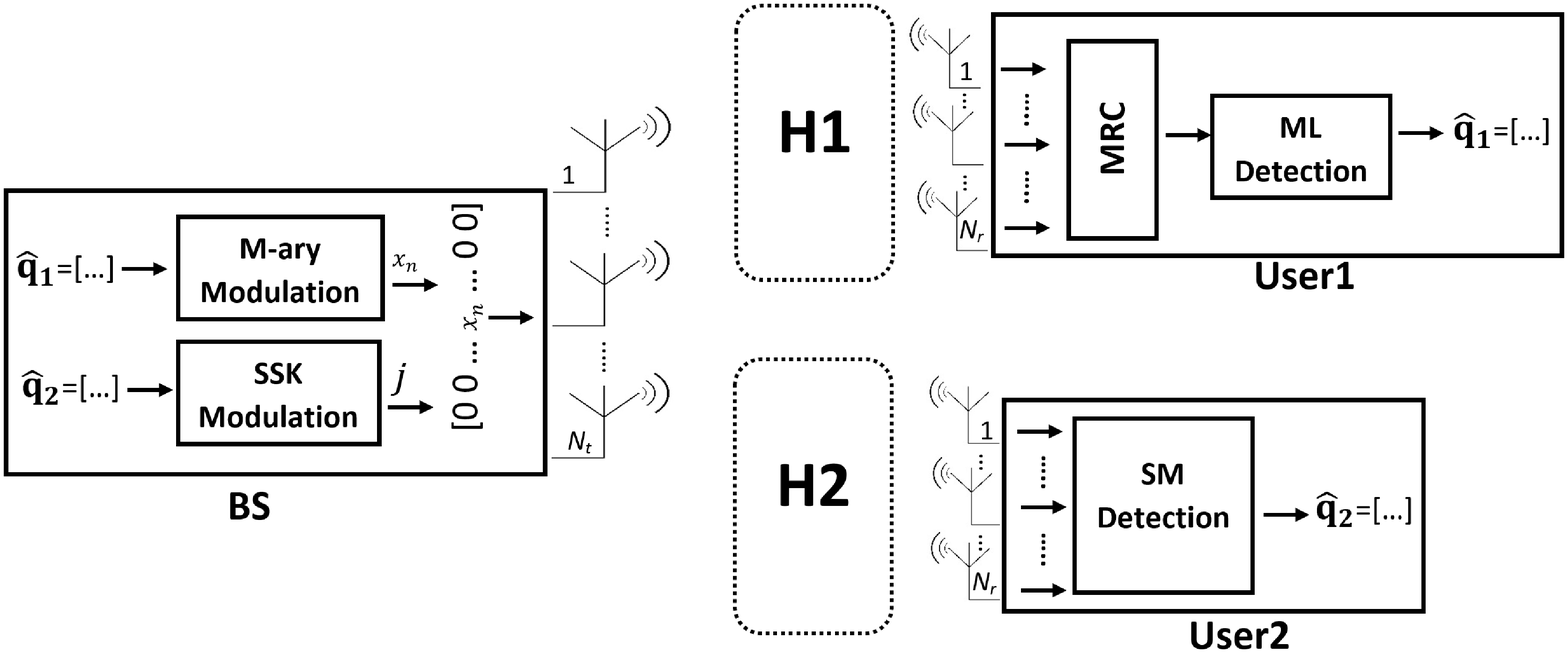}
    \caption{The illustration of the SMA}
    \label{fig1}
\end{figure}
The channel gain for the UE-1 and UE-2 are represented as $\mathbf{H1}$ and $\mathbf{H2}$, respectively.\footnote{In the following of this paper the notation used are as follows: the bold capital letters $\mathbf{'H'}$ show matrices, the lower case bold letters $\mathbf{'x'}$ show the vectors. We use $(.)^T$ for transpose, $(.)^H$ for conjugate transpose and $\left||.|\right|_F$ for the Frobenius form of a matrix/vector. We use $\left|.\right|$ for the absolute value of a scalar and $\binom{.}{.}$ for the binomial coefficient. $CN(\mu,\sigma)$ is a complex Gaussian distirubiton which has independent real and imaginary random variables with the $\mu$ mean and the $\frac{\sigma}{2}$ variance.} However, for the notation simplicity the user number is dropped for $\mathbf{H}$ and for the related vectors $\mathbf{h}$ in the rest of the paper.  The channel gains between each transmitting and each receiving antenna for a user are assumed to be flat fading and independent-identical distributed (i.i.d) as $CN(0,\sigma^2)$. The CSI is assumed to be known at the receivers. $\mathbf{q_1}$ and $\mathbf{q_2}$ are binary vectors of UE-1 and UE-2 with the $m_1=log_2(M)$ and $m_2=log_2(N_t)$ bits. $\mathbf{q_1}$ and $\mathbf{q_2}$ vectors are mutually mapped into another vector $\mathbf{x}$ with the size of $N_t$ in which only one element is different from zero. The non-zero element is obtained from the $M$-ary modulation constellation for the  $\mathbf{q_1}$ vector. The index of the non-zero element where to place is determined by the SSK modulation of the $\mathbf{q_2}$ vector. The resulting vector $\mathbf{x}$ is 
\begin{equation}
\begin{split}
\ \mathbf{x}=&\left[0 \ 0...x_n...0 \ 0\right]^T,  \quad  j=f_{SSK}(\mathbf{q_2}), x_n=f_{M-ary}(\mathbf{q_1}), \\
\ &\ \ \ \ \ \ \ \uparrow j^{th}position   
\end{split}
\end{equation}
where $f_{SSK}(.)$ and $f_{M-ary}(.)$ show the SSK and $M$-ary modulation mapping operations, respectively.  The $\mathbf{x}$ vector is transmitted to each user over MIMO channel $\mathbf{H}$. The MIMO channel $\mathbf{H}$ can be written in the form of vectors for each transmitting antenna $v$ as follows
\begin{equation}
\mathbf{H=\left[h_1,\ h_2, ...,\ h_{N_t}\right]},
\end{equation}
where
\begin{equation}
 \mathbf{h_v}=\left[ h_{v,1},\ h_{v,1}, ...,\ h_{v,N_r} \right]^T.
\end{equation}
The received vector for each user is given by $\mathbf{y_i}=\mathbf{h_{(v=j)}}x_{n}+{w_i}, i=1,2.$ where $w_i$ is the $N_r$-dim additive white Gaussian noise vector and each dimension is distributed as $CN(0,N_0)$.
\subsection{Detection at the users}
\subsubsection{UE-1}
The symbol of the UE-1 is sent according to the $M$-ary modulation constellation from the selected transmitting antenna and is received by $N_r$ receiving antennas. The transmitting antenna has no effect on the detection of the symbols of UE-1. Hence, the UE-1 implements a maximum likelihood (ML) receiver for  $M$-ary constellation with a maximum-ratio combining (MRC) as in the conventional OMA systems. The ML decision for the symbols of the UE-1 is 
\begin{equation}
\hat{x}_n=\argmin_{n}{\left||\mathbf{{y_1}-{h_{v=j}}}x_n|\right|^2}, \quad n=1,2,..M,
\end{equation}
where $x_n $ is the complex signal at the constellation point $n$ of the $M$-ary modulation.
\subsubsection{UE-2}
The binary symbols of the UE-2 are mapped into transmitting antenna index. Hence, the UE-2 must detect from which antenna the complex symbol of the UE-1 is sent.  Since the sent symbol from the active antenna is complex, we should implement an optimum SM detection algorithm in [12] instead of SSK detection [6].  The ML based SM detection is given
\begin{equation}
\left[\hat{j}, \hat{x}_n\right]=\argmin_{j,n}{\sqrt{\rho}\left||\mathbf{g_{j,n}}|\right|_F^2-2Re\{\mathbf{{y_2}}^H\mathbf{g_{j,n}\}}},  
\end{equation}
where $j=1,2,..N_t, n=1,2,..M$, $\mathbf{g_{jn}={h_{j}}}x_n$ and $\rho$ is the average signal-to-noise ratio (SNR) for each antenna. Although optimal SM detection detects the transmitting antenna number and the symbol of UE-1 mutually, UE-2 only takes the transmitting antenna number ($\hat{j}$) as output. So that, the symbol of the UE-2 is estimated. 
\section{Performance Analyses}
\subsection{Average Bit Error Probability (ABEP)}
\subsubsection{UE-1}
The conditional bit error probability of the UE-1 is equal to error probability of the well-known $1\times N_r$ SIMO system using the MRC. Hence, the conditional BEP for UE-1 is
$P_1(\left.e\right|_{h_j})=\alpha Q(\sqrt{\beta{\gamma_{1_b}}})$\  where ${\gamma_{1_b}}$ is the total received SNR per bit at the output of the MRC for UE-1. $\alpha$ and $\beta$  coefficients depend on the $M$-ary constellation. For example, for QPSK $\alpha=1$ and $\beta=2$. The ABEP of the UE-1 is obtained by averaging conditional BEP over instantaneous SNR $\gamma_{1_b}$ and becomes
\begin{equation}
\ P_1(e)=\int_{0}^{\infty}{P_1(\left.e\right|_{h_j})p_{\gamma_{1_b}}(\gamma_{1_b})d{\gamma_{1_b}}},
\end{equation}
where $p_{\gamma_{1_b}}(\gamma_{1_b})$ is the probability density function of $\gamma_{1_b}$ and in case of  ${h_{j,l}}$ is Rayleigh distributed, it is chi-square distributed with the $2N_r$ degree of freedom and given in [13] by
\begin{equation}
\ p_{\gamma_{1_b}}(\gamma_{1_b})=\frac{{\gamma_{1_b}}^{N_r-1}e^{\sfrac{-{\gamma_{1_b}}}{\overline{\gamma}_{1_b}}}}{\Gamma(N_r){\overline{\gamma}_{1_b}}^{N_r}},\ \ \ \  \ \ \ \overline{\gamma}_{1_b}=\sfrac{\rho\sigma_1^2}{\log_2{M}}.
\end{equation}

The closed-form expressions for the ABEP is obtained by substituting (7) into (6). For different modulation schemes, the ABEP expressions are provided in [14]. For BPSK/QPSK (gray coded) modulation is given as
\begin{equation}
\ P_1(e)=\left(\frac{1-\mu_1}{2}\right)^{N_r}\sum_{k=0}^{\eta}\binom{\eta+k}{k}\left(\frac{1+\mu_1}{2}\right)^k,
\end{equation}
where $\mu_1=\sqrt{\frac{\overline{\gamma}_{1_b}}{1+\overline{\gamma}_{1_b}}}$ and $\eta \triangleq N_r-1$.
\subsubsection{UE-2}The exact ABEP for the UE-2 cannot be determined, so that the union bound which is used ABEP analysis of SM/SSK systems in the literature widely, is analyzed. The union bound for ABEP of the optimal SM detection is given in [12] as
\begin{equation}
\ P(e)\le\sum_{j=1}^{N_t}\sum_{\hat{j}=1}^{N_t}\sum_{n=1}^{M}\sum_{\hat{n}=1}^{M}\frac{N(n,\hat{n})P(x_{j,n}\rightarrow x_{\hat{j},\hat{n}})}{MN_t},
\end{equation}
where $N(n,\hat{n})$ is the number of different bits between the symbols $x_n$ and $x_{\hat{n}}$. $x_{j,n}$ represent the symbol $x_{n}$ sent by the transmitting antenna $j$. $P(x_{j,n}\rightarrow x_{\hat{j},\hat{n}})$ is the pairwise error probability of ML decision given in (5) as $x_{\hat{j},\hat{n}}$ is estimated whereas $x_{j,n}$ is sent. At the UE-2, the output vector consists of only the estimated antenna vector bits, therefore the union bound for the UE-2 is determined as
\begin{equation}
\ P(e)\le\sum_{j=1}^{N_t}\sum_{\hat{j}}^{N_t}\frac{P(x_{j,n}\rightarrow x_{\hat{j},\hat{n}})}{N_t}.
\end{equation}
In case Rayleigh fading channels, by utilizing PEP given in [12] \footnote{In [12], PEP is only given for real modulation constellations (i.e., BPSK) }, for M-ary contellations the PEP is determined as
\begin{equation}
\begin{split}
\ P(x_{j,n}\rightarrow x_{\hat{j},\hat{n}})={\mu_2}^{N_r}\log_2{M} \sum_{k=0}^{\eta}\binom{\eta+k}{k}\left(1-\mu_2\right)^k,
\end{split}
\end{equation}
where $\mu_{2}=\frac{1}{2}\left(1-\sqrt{\frac{\sigma_a^2}{{1+\sigma}_a^2}}\right)$ and $\sigma_a^2=\frac{\rho{\sigma_2}^2\left(\left|x_{n}\right|^2+\left|x_{\hat{n}}\right|^2\right)}{4}$. By substituting (11) into (10), the union bound for the ABEP of the UE-2 turns out to be
\begin{equation}
\ P_2(e)\le { N_t {\mu_2}^{N_r}\log_2{M}}\sum_{k=0}^{\eta}\binom{\eta+k}{k}\left(1-\mu_2\right)^k.
\end{equation}
\subsection{Ergodic Sum Rate}
The achievable (Shannon) capacities of the users for the proposed SMA system are
\begin{equation}
\ {R_1}^{SMA}=\log_2{(1+{\gamma_1})}, \ \ {R_2}^{SMA}=\log_2{(N_t)}.
\end{equation}
where $\gamma_1=\gamma_{1_b} log_2{M}$. The achievable capacity of the UE-2 only depends on the number of the transmitting antennas (when the receiver sensitivity is ignored). Hence, to obtain ergodic sum rate of the system, ergodic capacity of the UE-1 should be analyzed. The ergodic capacity of the UE-1 is given 
\begin{equation}
{\overline{C}}_1=\int_{0}^{\infty} \log_2{(1+{\gamma_1})}p_{\gamma_1}(\gamma_1)d\gamma_1
\end{equation}
After substituting PDF given in (7) into (14), with some algebraic manipulations ergodic capacity of the UE-1 is obtained by utilizing [15, eq. (4.333.5)]. Ergodic sum rate is given as ${\overline{C}}={\overline{C}}_1+{R_2}^{SMA}$ and is derived
\begin{equation}
{\overline{C}}=\log_2{(N_t)}+\frac{\log_2{e}}{\Gamma(N_r)}\sum_{k=0}^{\eta}\frac{\eta !}{(\eta-k)!} \left[\frac{(-1)^{\eta-k-1}}{\rho^{\eta-k}}e^{\sfrac{1}{\rho}}E_i\left(-\frac{1}{\rho}\right)+\sum_{j=1}^{\eta-k}\frac{(j-1)!}{\left(-\rho\right)^{\eta-k-j}}\right]
\end{equation}
where $E_i(.)$ and $\Gamma(.)$ are the exponential integral and the gamma function, respectively. Once UE-1 and UE-2 are determined as the near user and far user for NOMA, respectively, the achievable rate of NOMA users are given as ${R_1}^{NOMA}=\log_2\left(1+a_1{\gamma_1}\right)$ and ${R_2}^{NOMA}=log_2\left(1+\sfrac{a_2{\gamma_2}}{a_1{\gamma_2}+1}\right)$ [2]. Where $a_1$ and $a_2$ are the power allocation (PA) coefficients for the users. $a_1=1-a_2$  and $a_2>a_1$. By the placement of large number of the transmitting antenna for UE-2, it can easily seen that for the all PA coefficients
\begin{equation}
{R_1}^{SMA}>{R_1}^{NOMA},\ {R_2}^{SMA}>{R_2}^{NOMA}.
\end{equation}
\subsection{Outage Probability}
The outage probabilities of the users are 
\begin{equation}
P_i(out)=P\left({{R_i}}^{SMA}<{\acute{R}}_i\right), \ \ \ i=1,2 \\
\end{equation}
where ${\acute{R}}_i, i=1,2$ are the targeted data rates of the users.  For the UE-1
\begin{equation}
\begin{split}
P_1(out)&=P\left(\gamma_1<2^{{\acute{R}}_1}-1\right) \\
&=F_{\gamma_1}\left(2^{{\acute{R}}_1}-1\right).
\end{split}
\end{equation}
$F_{\gamma_1}(.)$ is the cumulative density function, and for the Rayleigh fading channel it is given [13] as
\begin{equation}
F_{\gamma_1}(\theta)=1-{e^{\sfrac{-\theta}{\overline{\gamma}_1}}\sum_{k=1}^{N_r}\frac{\left(\sfrac{\theta}{\overline{\gamma}_1}\right)^{k-1}}{\left(k-1\right)!}}.
\end{equation}
The outage probability of the UE-1 is obtained by substituting $\theta=2^{{\acute{R}}_1}-1$ into (19),

For the UE-2, the outage event does not occur when the targeted data rate is less than the number of bits can be mapped into the transmitting antennas (i.e., ${\acute{R}}_2<{log}_2(N_t)$). In this case, it becomes $P_2(out)=0$ so that the SMA outperforms to the NOMA under the same targeted rate. 
\section{Simulation Results}
In this section, validation of the derived expressions in the previous section are provided via computer simulations. In addition,  to show superiority of the SMA system, the comparison with the NOMA systems are provided in terms of the all performance metrics (i.e., bit error rate, outage and ergodic sum rate). In all figures, average channel gain between each transmitting and receiving antenna is assumed to be equal to $1$ (i.e., $\sigma_1^2=\sigma_2^2=1$). It is chosen as $N_r=N_t$ in the SMA systems and , the modulation level of the UE-1 in SMA and of the both users in NOMA systems are chosen equal to $M=N_t$ for the fair comparison. 
The PA coefficients for NOMA users are chosen as $a_1=0.2$ and $a_2=0.8$ as given in [2], [3]. In all figures, simulations are provided for $10^8$ channel realizations. 

In Fig. 2, the BER comparison of the SMA and NOMA systems are presented respect to the average transmitted SNR. 
SMA outperforms substantially to the NOMA systems. The full diversity order (i.e., $N_r$) is achieved for the both users in SMA. 
\begin{figure}[!t]
    \centering
    \includegraphics[width=8.5cm,height=5.4cm]{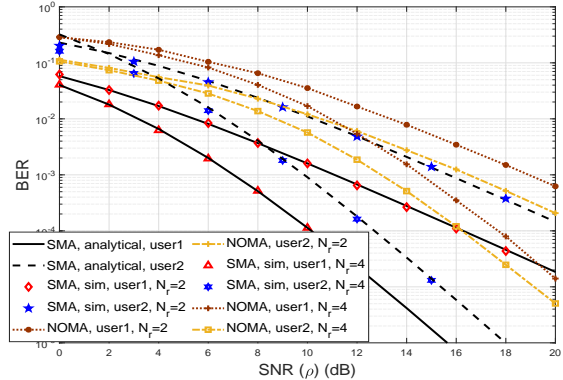}
    \caption{BER Comparison of SMA and NOMA }
    \label{fig2}
\end{figure}
In Fig 3, the ergodic sum rate comparison of the SMA and SIMO-NOMA systems are provided. SMA systems can achieve higher sum rate than NOMA for all  number of receiving antennas ($N_r$). In addition, achievable rate of UE-2 in SMA can be easily improved by increasing the number of the transmitting antennas ($N_t$) so that the sum rate of SMA will be improved.
\begin{figure}[!t]
    \centering
    \includegraphics[width=8.5cm, height=5.4cm]{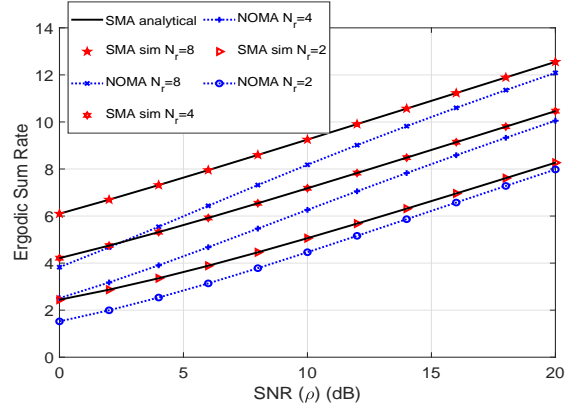}
    \caption{Sum Rate Comparison of SMA and NOMA}
    \label{fig3}
\end{figure}
Lastly, the outage comparison of the SMA and NOMA systems are given in Fig4. The targeted data rates of the users are chosen according to the number of the transmitting antenna of SMA (i.e., $\acute{R}_1=\acute{R}_2=log_2(N_t)$). The full diversity order is achieved for the outage performance of the UE-1 in SMA systems. The outage performance of the UE-2 in SMA is not provided due to $P_2(out)=0$ (when receiver sensitivity is ignored). SMA is superior to NOMA systems in terms of the outage performance as well. It is worth pointing out that provided simulations results of SMA match well with the derived analytical expressions in (8), (12), (15) and (19). 
\begin{figure}[!t]
    \centering
    \includegraphics[width=8.5cm,height=5.4cm]{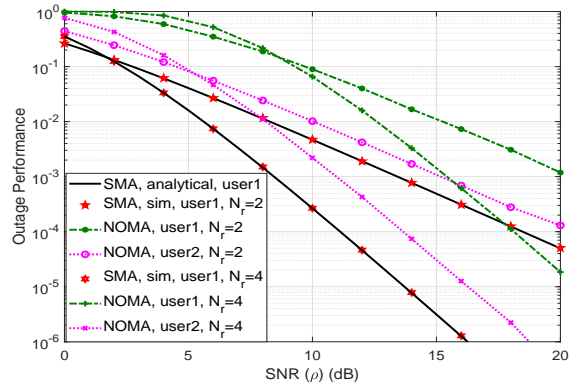}
    \caption{Outage Comparison of SMA and NOMA}
    \label{fig4}
\end{figure}
\section{Conclusion}
In this letter, performances of the spatial multiple access (SMA) proposed as an alternative of NOMA to deal with the drawback of the NOMA systems caused by the inter-user interferences, are investigated. The analytical ABEP, ergodic sum rate and the outage probability expressions are derived. The comparison of the SMA and NOMA systems for all performance metrics (i.e., bit error rate,ergodic sum rate and outage) are simulated. The results reveal that 1) SMA is superior to NOMA for all three metrics. 2) The full diversity order (number of receiving antennas) is achieved for the SMA system. 3) SMA consumes much less power than NOMA to meet the same performance which is very promising for the energy efficiency. 4) SMA has much less complexity than NOMA since the SIC implementation at the receiver and the channel ordering with the power allocation algorithms at the transmitter are no longer have to be succeeded besides only one RF chain at the transmitter is needed for SMA. Lastly, the proposed SMA system can be expanded for the applications with the NOMA systems to meet higher number of users by achieving better performance metrics.

%
\ifCLASSOPTIONcaptionsoff
  \newpage
\fi

\begin{thebibliography}{1}
\bibitem{1} Y. Liu \textit{et al.}, ``Non-Orthogonal Multiple Access for 5G'', \textit{Proceedings of IEEE}, 2017, no. December, pp. 2017-2018.
\bibitem{2}Z. Ding \textit{et al.}, ``On the Performance of Non-Orthogonal Multiple Access in 5G Systems with Randomly Deployed Users'', \textit{IEEE Signal Process. Lett.}, vol. 21, no. 12, pp. 1501-1505, 2014.
\bibitem{3}	Y. Saito \textit{et al.}, ``System-level performance evaluation of downlink non-orthogonal multiple access (NOMA)'', \textit{IEEE Int. Symp. Pers. Indoor Mob. Radio Commun. PIMRC}, vol. 2, pp. 611-615, 2013.
\bibitem{4}F. Kara and H. Kaya, ``BER performances of downlink and uplink NOMA in the presence of SIC errors over fading channels'', \textit{IET Commun.}, vol. 12, no. 15, pp. 1834-1844, 2018.
\bibitem{5}	R. Mesleh \textit{et al.}, ``Spatial modulation - A new low complexity spectral efficiency enhancing technique,” \textit{First Int. Conf. on Commun. and Netw. in China, ChinaCom ’06}, 2007.
\bibitem{6}J. Jeganathan \textit{et al.}, ``Space shift keying modulation for MIMO channels'', \textit{IEEE Trans. Wirel. Commun.}, vol. 8, no. 7, pp. 3692-3703, 2009.
\bibitem{7}	N. Serafimovski \textit{et al.}, ``Multiple access spatial modulation'', \textit{EURASIP J. Wirel. Commun. Netw.}, vol. 2012, no. 1, p. 299, 2012.
\bibitem{8} S. Narayanan \textit{et al.}, ``Multi-user spatial modulation MIMO'', \textit{IEEE Wirel. Commun. Netw. Conf. WCNC}, vol. 1, no. 3, pp. 671-676, 2014.
\bibitem{9} X. Wang \textit{et al.}, ``Spectral Efficiency Analysis for Downlink NOMA Aided Spatial Modulation with Finite Alphabet Inputs'',\textit{ IEEE Trans. Veh. Technol.}, vol. 66, no. 11, pp. 10562-10566, 2017.
\bibitem{10} Y. Chen \textit{et al.}, ``Performance analysis of NOMA-SM in vehicle-to-vehicle massive MIMO channels'', \textit{IEEE J. Sel. Areas Commun.}, vol. 35, no. 12, pp. 2653-2666, 2017.
\bibitem{11} C. Zhong \textit{et al.}, ``Spatial Modulation Assisted Multi-Antenna Non-Orthogonal Multiple Access'', \textit{IEEE Wirel. Commun.}, vol. 25, no. 2, pp. 61-67, 2018.
\bibitem{12}	J. Jeganathan, A. Ghrayeb, and L. Szczecinski, ``Spatial modulation: optimal detection and performance analysis'', \textit{IEEE Commun. Lett.}, vol. 12, no. 8, pp. 545-547, 2008.
\bibitem{13}	J. G. Proakis, \textit{Digital Communications}, 5th ed., New York: McGraw-Hill, 2008, p. 1150.
\bibitem{14}	M. S. Alouini, ``A unified approach for calculating error rates of linearly modulated signals over generalized fading channels'', \textit{IEEE Trans. Commun.}, vol. 47, no. 9, pp. 1324-1334, 1999.
\bibitem{15}I. S. Gradshteyn and I. M. Ryzhik, \textit{Table of Integrals, Series, and Products}, 5th ed. San Diego: CA: Academic Press, 1994.
\end{thebibliography}
\end{document}